\documentclass[12pt]{article}
\title{\LARGE\bf Maximal entropy random networks with given degree distribution}
\date{}
\author{}
\begin{document}
\maketitle

\vspace{-1.2cm}

\centerline{\large Michel Bauer\footnote[1]{Email:
    bauer@spht.saclay.cea.fr} and Denis Bernard\footnote[2]{Member of
    the CNRS; Email : dbernard@spht.saclay.cea.fr}}

\vspace{.3cm}

\centerline{\large Service de Physique Th\'eorique,}
\centerline{CEA/DSM/SPhT, Unit\'e de recherche associ\'ee au CNRS}
\centerline{CEA-Saclay, 91191 Gif-sur-Yvette cedex, France} 


\vspace{.3cm}

\begin{abstract}

Using a maximum entropy principle to assign a statistical weight to
any graph, we introduce a model of random graphs with arbitrary degree
distribution in the framework of standard statistical mechanics. We
compute the free energy and the distribution of connected components. 
We determine the size of the percolation cluster above the
percolation threshold. The conditional degree distribution on the
percolation cluster is also given.  We briefly present the analogous
discussion for oriented graphs, giving for example the percolation
criterion.

\end{abstract}

\section{Introduction}
\label{sec:intro}

The statistical properties of networks, either biological, social or
technological, have received a lot of attention recently both
experimentally and theoretically, See eg. refs.\cite{dorogotsev,barab}.
 One of the most studied features of
those networks is the degree distribution, which describes 
the probability for the vertices to have $0,1,\cdots$ neighbors. 
One striking observation is that, in many examples, the degree
distribution is large so that the probability to have $n$ neighbors
decreases slowly with $n$. Several models (static or evolving)
predict such a behavior. More generally, they contain enough tunable
parameters to reproduce almost any degree distribution.

However, the static models are in general not conveniently defined
within the language of statistical mechanics (see
ref.\cite{dorogotsev}, which motivated our interest in this question).
This is for instance the case with the most intuitive proposal
\cite{newm}: generate independently half edges for each vertex, with
the appropriate distribution, and then join the half edges at random.
This makes it rather easy to generate random graphs, but does not
assign in a simple way a probability to any given simple graph :
it is formally complicated to eliminate multiple edges.  Another
proposal made in \cite{kry} has some formal technical similarity with
our work but really leads to a different model.

It is moreover obvious, if not always apparent in the literature,
that the knowledge of the degree distribution leaves many 
statistical properties of the graphs undetermined, even if one insists
that all vertices are equivalent. This arbitrariness is a problem,
because most of the time the models used to fit the behavior of say a
communication network are just ingenious constructions~: they are not
derived from clear basic principles. Such principles may be out of our
reach at the moment, and so is a classification of all random graph
models with certain apriori properties. Consequently, we propose to
use maximum entropy as a criterion to build a model that does not make
any a priori bias, incorporating what we know -- in this case the degree
distribution -- but nothing else. Comparison with real networks is a way
to get evidence for other striking features that might be overlooked
today. 

The maximal entropy principle is applied  here to deal with constraints
on the degree distribution but it can clearly be engineered
to deal with other constraints.

This paper is organized as follows :

-- Section 2 starts with the main definitions, goes on with a quick
reminder on the Molloy--Reed model \cite{molloy} 
and continues with the definition of
the maximum entropy model. We use it to reformulate the standard
Erd\"os-Renyi random graph model. Then we derive a few general
identities valid for the maximal entropy model, and study the
distribution of connected components. 
Our model is a close cousin of the Molloy--Reed model 
and we make the connexion precise below.
Finally we discuss the possibility of numerical simulations. 

-- Section 3 studies the thermodynamical limit when the number $N$ of
sites is large, but the number of edges scales like $N$, hence the
name finite connectivity limit for this regime. We derive the
equations that determine all physical quantities in this regime : free
energy, distribution of the number of edges incident at a vertex, ...
We then study the connected components, derive the criterion for the
existence of a percolation cluster and the formula for its size.
Finally, we study the distribution of the number of edges incident at
a vertex in the percolation cluster.

-- Section 4 analyzes the generalization to oriented graphs, ending
with the criterion for the existence of a percolation cluster and the
formula for its size.

\section{The model}
\label{sec:model}

\subsection{General definitions}
\label{sec:defs}

In the following, we shall concentrate on labeled simple unoriented
graphs, or equivalently on symmetric $0-1$ matrices, with vanishing
main diagonal : the matrix element $(i,j)$ is $1$ if vertices $i$ and
$j$ are connected by an edge and $0$ else. So we use the same letter
$G$ to denote the graph and its adjacency matrix with matrix elements
$G_{i,j}$. In the sequel, unless otherwise stated, the term graph
refers to labeled simple unoriented graph. The number of edges of a
graph $G$ is denoted by $E(G)$ and the number of vertices by $V(G)$.

The row-sum $\hat{G}_i=\sum_j G_{i,j}$ is the number of neighbors of
site $i$. The degree distribution of $G$ is the sequence
$\tilde{G}_k=\# \{i\ \mbox{such that } \hat{G}_i=k \}$, so that
$\tilde{G}_0$ is the number of isolated points of $G$, $\tilde{G}_1$
is the number of vertices of $G$ with exactly one neighbor, and so on.

Not every integer sequence can appear as the degree distribution of a
graph on $N$ vertices : $\tilde{G}_k=0$ for $k\geq N$, $\sum_k
\tilde{G}_k=N$ and $\sum_k k \tilde{G}_k$ is even because this number
counts twice the number of edges of $G$, i.e. $\sum_{i,j} G_{i,j}$.
There are other less obvious constraints. We call the sequences that
appear as degree distribution of a graph on $N$ vertices
$N$-admissible. There is a relatively simple family of inequalities
that characterizes $N$-admissible sequences, but for instance the
(asymptotic) counting of $N$-admissible sequences is still unknown.

\subsection{The Molloy--Reed model}
\label{sec:molloyreedmod}

Before we introduce our model, let us describe the method of Molloy
and Reed \cite{molloy} which can be interpreted as a kind of
microcanonical version of our model. The idea is quite elegant : for
any integer $N$ fix an $N$-admissible sequence $\{m_{N,k}\}_{k\geq 0}$
and take as probability space the set ${\mathcal G}_{\{m_{N,k}\}}$ of
graphs with degree distribution $\{m_{N,k}\}_{k\geq 0}$, endowed with
the uniform (counting) probability. By construction, in ${\mathcal
  G}_{\{m_{N,k}\}}$, the probability that vertex $i \in [1,N]$ has $k$
neighbors is exactly $m_{N,k}/N$.

Molloy and Reed show that if the sequence $\{m_{N,k}/N\}$ converges
(uniformly) to a probability distribution $\{\pi _k\}$,($\sum _k \pi
_k=1$), under one technical assumption, the space ${\mathcal
  G}_{\{m_{N,k}\}}$ converges in an appropriate sense to a random
graph ensemble ${\mathcal G}_{\{\pi _k\}}$ on which standard questions
can be formulated and answered :

-- the probability in ${\mathcal G}_{\{\pi_k\}}$ that a given vertex
has $k$ neighbors is -- not surprisingly -- $\pi _k$,

-- Molloy and Reed give a criterion for the presence or absence of a
giant component.

Heuristic arguments \cite{newm} show that the `intuitive' model (which
does not in general lead to simple graphs) namely ``generate
independently half edges for each vertex, with distribution $\{\pi
_k\}$ and then join the half edges at random'', has the same
thermodynamical -- large $N$ -- properties as the Molloy--Reed model.

\subsection{The maximum entropy model}
\label{sec:maxentmod}

To start with, we fix an integer $N \geq 1$, and a probability
distribution $\{\pi_{N,k}\}$ ($\sum_{k=1}^{N-1}\pi_{N,k}=1$). We want
to look for a probability distribution $\{p_G\}$ on the set of graphs
on $N$ vertices such that for any vertex $i$, $\sum_{G;\,
  \hat{G}_i=k}p_G =\pi_{N,k}$ where, here and below, the notation
means that the sum is restricted to graphs such that $\hat G_i$, the
number of neighbors of vertex $i$ in $G$, equals $k$. With words, we
look for a probability distribution $\{p_G\}$ on the set of graphs on
$N$ vertices such that the probability that vertex $i$ has $k$
neighbors is $\pi_{N,k}$. As explained in the introduction, this
requirement is far from fixing the probability distribution.

We also want
this probability distribution to have no other bias. So we look for a
distribution $\{p_G\}$ with maximal entropy
\footnote{Notice that, if no constraint is imposed, the
uniform counting measure has  maximal entropy. 
This measure can be described as follows: 
the probability of an edge between
vertices $i$ and $j$ is $1/2$ independently of the presence or absence
of any other edge.}.  

Hence  we want to maximize $\sum_G p_G \log p_G$ under the constraints
$$\sum_{G ;\, \hat{G}_i=k}p_G
=\pi_{N,k}$$ which we implement as Lagrange multipliers. The extremum
conditions for
$$\sum_G p_G (\log p_G +1) - \lambda (\sum_G p_G -1) - \sum_{i,k}
\lambda_{i,k}(\sum_{G ;\, \hat{G}_i=k}p_G-\pi_{N,k})$$
are
$$p_G=e^{\lambda+ \sum_i \lambda_{i,\hat{G}_i}} \mbox{, } \sum_G p_G
=1\mbox{ and } \sum_{G ;\, \hat{G}_i=k}p_G
=\pi_{N,k}.$$

It is not obvious to us that these equations always have a
solution and that this solution is unique and symmetric i.e.
$\lambda_{i,k}$ does not depend on $i$  \footnote{In the large $N$
limit, some spontaneous symmetry breaking might occur. We shall not
pursue this questions here.}. But as usual in statistical mechanics, we
can reverse the logic : we start from an arbitrary sequence of positive
numbers $t_k=e^{\lambda_k}$ and define $p_G$:
\begin{equation}
p_G\equiv e^{\lambda + \sum_i \lambda_{\hat{G}_i}} =
e^{\lambda}\,\prod _k t_k^{\tilde{G}_k}  \label{defpg}
\end{equation}
with suitably adjusted $\lambda$ so as to ensure $\sum_G p_G =1$. 

We define the weight $w_G$ of a graph as 
$$
w_G \equiv \prod _k t_k^{\tilde{G}_k}=\prod _i t_{\hat{G}_i}
$$ 
and the partition  function as the sum of weights
$$Z_N \equiv \sum_G w_G =\sum_G \prod _k t_k^{\tilde{G}_k}.$$
Hence $\lambda=-\log Z_N$ and $p_G= w_G/Z_N$. 

By construction the probability distribution $\pi_{N,k}$
that vertex $i$ has $k$ neighbors is $i$-independent. 
Recall that $\tilde{G}_k$ is the number of vertices 
with $k$ neighbors in $G$ so that 
$$ 
N\, \pi_{N,k}=
\sum _i \sum_{G ;\,  \hat{G}_i=k}p_G = \sum _G
\tilde{G}_k p_G=\frac{1}{Z_N}\sum _G \tilde{G}_k w_G.$$
But $\tilde{G}_k w_G= \partial w_G/\partial \lambda_k$ so 
$$ \pi_{N,k} =
\frac{1}{N}\frac{\partial \log Z_N}{\partial\lambda_k}
=\frac{1}{N}t_k\frac{\partial \log Z_N}{\partial t_k}.$$

\subsection{The Erd\"os--Renyi model revisited}
\label{sec:ERrev}

As a first application, but also as a preparation to section
\ref{sec:fincon}, let us reinterpret the standard Erd\"os-Renyi random
graph model \cite{erdos} in our framework. Recall that in the
Erd\"os-Renyi model, the edges are 
described by independent binomial variables, each edge being drawn
with probability $p$. Recall that $E(G)$ denotes the number of edges
of a graph $G$.  The probability of the graph $G$ is simply
$p^{E(G)}(1-p)^{N(N-1)/2-E(G)}$ which we rewrite as $(1-p)^{N(N-1)/2}
\left(\frac{p}{1-p}\right)^{E(G)}$.  Now $2E(G)=\sum_k k \tilde{G}_k$.
So letting $t_k\equiv \left(\frac{p}{1-p}\right)^{k/2}$ shows that the
Erd\"os-Renyi model is the maximal entropy model such that the
probability that a vertex has $k$ neighbors is ${N-1 \choose
  k}p^k(1-p)^{N-1-k}$. The average number of neighbors is $(N-1)p$. In
the large $N$ limit, an interesting regime occurs when this number is
kept fixed, so that $p \sim \alpha /N$ and $\alpha$ is the control
parameter. The important observation is that if $p$ scales like
$N^{-1}$, the parameters $t_k$ scale like $N^{-k/2}$. In section
\ref{sec:fincon}, we shall see that indeed generically these scaling
relations ensure that $\log Z_N$ scales like $N$, as any ``good'' free
energy should.

\subsection{Useful relations}
\label{sec:userel}

We establish a few formul\ae\ which will be central in the following
discussion.

The sequence  $\{Z_N\}_{N \geq 1}$ satisfies a first order functional
recursion relation that will prove useful in the subsequent analysis.
We define the formal Laurent series $H(\omega,t_0,\cdots,t_k,\cdots)= 
\sum_k t_k \omega ^{-k}$.

Suppose $N\geq 2$. Then $Z_N$ is the constant (i.e. of degree $0$)
term in the $\omega$-expansion of the product 
$$ H(\omega,t_0,\cdots,t_k,\cdots) Z_{N-1}(t_0+\omega t_1,\cdots,t_j+\omega
t_{j+1},\cdots).$$
The $\omega$-expansion of the product is well-defined because both
factors involve at most a finite number of terms of positive degree. 

The proof of this relation goes as follows. If $G'$ is a graph on $N-1$
vertices $1,\cdots,N-1$, it can be completed to a graph $G$ on $N$
vertices in the following ways : add vertex $N$ and $k=0,\cdots,N-1$
edges emerging from $N$. Attach these edges to any $k$ distinct
vertices of $G'$. There is a simple relation between the weights of
$G$ and $G'$ because one vertex of degree $k$ has been added (this
is taken care of by the term $t_k$ in $H$), and $k$ vertices in $1,\cdots,N-1$
have seen their degree increased by $1$ so, in
$w_{G'}=\prod_{1}^{N-1}t_{\hat{G}'_i}$, $k$ of the factors $t_j$ are
replaced by $t_{j+1}$ (this is taken care of by replacing all $t_j$'s in
$Z_{N-1}$ by $t_j+\omega t_{j+1}$ and expanding to order $k$ in
$\omega$). Note that the relation is also true for $N=1$ if we make
the natural choice $Z_0=1$.

We rewrite this result as a (formal) contour integral
\footnote{The symbol $\oint$ denotes the contour integral
$\frac{1}{2i\pi}\int$ along small contour surrounding the origin.}~:
$$Z_N(t_0,\cdots,t_j,\cdots) = \oint \frac{d \omega}{\omega} \sum_k
t_k \omega ^{-k}Z_{N-1}(t_0+\omega t_1,\cdots,t_j+\omega
t_{j+1},\cdots).
$$
The same argument, based on enumerating the ways the point $N$ can be
 linked to the remaining part of the graph under the condition that 
$\hat G_N=k$, shows that 
$$Z_N \pi_{N,k} =t_k \oint \frac{d \omega}{\omega} \omega
^{-k}Z_{N-1}(t_0+\omega t_1,\cdots,t_j+\omega t_{j+1},\cdots),$$
to be compared with the formula of Section \ref{sec:maxentmod}.
These two formul\ae\  for $\pi_{N,k}$ are not so trivially
equivalent because they involve different rearrangements of the sum of
weights.

\subsection{Component distribution}
\label{sec:compdis}

We study the distribution of sizes of connected components. 

Define $W_n$ by 
$$ 
W_n = {\sum_{G;\, V(G)= n } }^{\hskip - .5 truecm c} 
\hskip  .4 truecm  w_G,
$$
where $\sum^c$ denotes the sum over connected graphs.
Observe that if $G$ splits as a disjoint union of two subgraphs $G_1$
and $G_2$ ($G$ contains no edge joining a vertex of $G_1$ to a vertex
of $G_2$), the weight of $G$ factorizes : $w_G=w_{G_1}w_{G_2}$. So the
total weight of graphs $G$ of size $N$ that are the disjoint union of
$k_1$ connected components of size $1$ (i.e. isolated points), $k_2$
connected components of size $2$,$\cdots$,$k_n$ connected components
of size $n$,$\cdots$ (so $\sum_{n \geq 1} nk_n=N$) is
$$ \frac{N!}{\prod_n k_n! n!^{k_n}} \prod _n W_n^{k_n}.$$
The combinatorial factor just counts the number of ways to split the
$N$ vertices of $G$ in packets of the right size. Summing over all
possible $k_n$'s gives back $Z_N$ :
$$ Z_N= \sum_{k_n\geq 0,\; \sum_{n \geq 1} nk_n=N} \frac{N!}{\prod_n
  k_n! n!^{k_n}} \prod _n W_n^{k_n}.$$
This formula allows to view
$Z_N$ not as a function of the $t_k$'s but as a function of the
$W_n$'s, and using this interpretation, we see that, denoting by
$C_m(G)$ the number of connected components of size $m$ in the graph
$G$, the average number of components of size $m$ in the random graph
model is
$$\sum_G p_G C_m(G) =
W_m\frac{\partial \log Z_N}{\partial W_m}.$$
So $m W_m\frac{\partial
  \log Z_N}{\partial W_m}$ is the average number of sites belonging to
components of size $m$, and summing over $m$ we should have $\sum _m m
W_m\frac{\partial \log Z_N}{\partial W_m}= N$.
This is simply the statement that $Z_N$ is a homogeneous
function of degree $N$ in the $W_n$'s if $W_n$ is assigned degree $n$.

This can be rephrased in compact form. Introduce a (complex or
formal)\footnote{In this section, some of the computations we make
require that the $t_k$'s satisfy some properties so as to ensure
that the series we write have a finite domain of convergence. For
instance, we could assume that only a finite (though arbitrarily
large) number of $t_k$'s are non vanishing. Alternatively we could
work with formal power series.} variable $z$ and define 
$ Z \equiv \sum_{N \geq 0} \frac{z^N}{N!}Z_N,$
the $z$-generating function for the $Z_N$'s. Replacing $Z_n$ by its
expression in terms of the $W_n$'s, we get the
(well-known) fact that $Z=e^W$
where $ W = \sum_{n \geq 1} \frac{z^n}{n!}W_n.$
Conversely, one retrieves $Z_N$ by 
\begin{equation}
 Z_N = N!\oint \frac{dz}{z}z^{-N} e^{\sum_{n \geq 1}
\frac{z^n}{n!}W_n}.
\label{superzn}
\end{equation}

The average number of components of
size $n$ in the random graph is thus
$$ W_n\frac{\partial \log Z_N}{\partial W_n}= 
\frac{N!}{n!(N-n)!}\, W_n\, \frac{Z_{N-n}}{Z_N}.$$

Similarly, the average number of times a given graph $g$ of
size $n\leq N$ appears as a connected component 
in the random graph $G$ of size $N$ is
$$
\frac{N!}{n!(N-n)!}\, w_g\, \frac{Z_{N-n}}{Z_N}.$$

\subsection{Discussion}
\label{sec:disc}

A crucial observation is that the weight of a graph depends
only on its degree distribution, as in the Molloy--Reed model. But
whereas in the Molloy--Reed model the weight of a graph is $0$ unless
it has the correct degree distribution, the degree distribution fluctuates in
our model. So our model is a canonical description of a random
graph model with given ``number of edges distribution at a vertex'', 
and the Molloy--Reed model a microcanonical one. 

That the two models turn out to be equivalent in some large $N$ limit
is maybe not surprising. However, note that contrary to standard
statistical mechanics (when only a few quantities, for instance energy
and number of particles, fluctuate in the canonical description but
are fixed in the microcanonical one) the constraint hypersurface of the
microcanonical model has a codimension that gets larger and larger as
$N$ grows.

Finally, let us observe that the maximum entropy model is well suited
for standard thermodynamical simulations, namely heatbath algorithms
or metroplolis algorithms. This is because contrary to the 
`naive' model or the Molloy--Reed model
 the phase space has a simple structure.

\section{Finite connectivity limit}
\label{sec:fincon}

\subsection{General analysis}
\label{sec:genal}

As suggested at the end of section \ref{sec:maxentmod} by the special
case of the Erd\"os--Renyi model, we shall show that a thermodynamic
limit occurs in the large $N$ limit if $t_k$ scales like $N^{-k/2}$.
Note that in this case $w_G$, the weight of $G$ scales like
$N^{-E(G)}$, where as before $E(G)$ stands for the number of edges of
$G$.

The starting point of the analysis will be the functional equation
established in section \ref{sec:userel} :
$$Z_N(t_0,\cdots,t_j,\cdots) = \oint \frac{d \omega}{\omega} \sum_k
t_k \omega ^{-k}Z_{N-1}(t_0+\omega t_1,\cdots,t_j+\omega
t_{j+1},\cdots).$$
We set $t_k=\tau _k N^{-k/2}$ and define $F_N(\tau
_.)\equiv \frac{1}{N} \log Z_N(t_.)$. Substituting $\omega N^{-1/2}$
for $\omega$ leads after a few manipulations to
$$ e^{NF_N(\tau _.)}=\oint \frac{d \omega}{\omega} \sum_k
\tau _k \omega ^{-k} e^{(N-1)F_{N-1}((\tau
_.+\frac{\omega}{N}\tau _{.+1})(1-1/N)^{k/2})}.$$

This equation still involves no approximation. Now we make the
usual thermodynamical hypothesis, namely that $NF_N(\tau _.)-
(N-1)F_{N-1}(\tau _.)$ has a limit, say $F(\tau _.)$ when $N
\rightarrow \infty$. This implies in particular that $F_N(\tau _.)$
converges to $F(\tau _.)$. The above equation has then a large $N$
limit. To see it clearly, we rewrite it as
$$
e^{NF_N(\tau _.)-(N-1)F_{N-1}(\tau _.)}=\oint \frac{d \omega}{\omega}
\sum_k \tau _k \omega ^{-k} e^{(N-1)[F_{N-1}((\tau
_.+\frac{\omega}{N}\tau _{.+1})(1-1/N)^{k/2})-F_{N-1}(\tau_.)]}.$$
In the large $N$ limit, this leads to the equation
\begin{equation}
1=\bar{y}\sum_k \tau _k \frac{ \bar{x}^k}{k!},
\label{1yvx}
\end{equation}
where we have defined 
$$\bar{y}\equiv e^{-F-\frac{1}{2}\sum_k k\tau
  _k\frac{\partial F}{\partial \tau _k}} \quad ,\quad 
\bar{x}\equiv\sum_k \tau
  _{k+1}\frac{\partial F}{\partial \tau _k}.
$$
  On can take an analogous limit of other relations in
\ref{sec:userel} to obtain the more detailed equations for
the degree distribution,
\begin{equation}
 \pi _k=\tau _k \frac{\partial F}{\partial \tau _k} =\bar{y} \tau
_k \frac{ \bar{x}^k}{k!}.
\label{pixtau}
\end{equation} 
Eq.(\ref{1yvx}) ensures that this
distribution is correctly normalized, $\sum_k\pi_k=1$.

 The parameter $\bar{x}$ posseses a simple interpretation.
We start from the relation 
$ \frac{\partial F}{\partial \tau _k}=\bar{y} \tau
_k \frac{ \bar{x}^k}{k!},$ multiply it by $\bar{x} \tau _{k+1}$ 
and sum over $k$ to get 
\begin{equation}
\bar{x}^2=\bar{y}\sum_k \tau _{k+1} \frac{
  \bar{x}^{k+1}}{k!}=\sum_k k \pi _k, \label{xbar}
\end{equation}
so that $\bar{x}^2$ is the first moment of the
distribution $\pi_k$ for the number of edges incident at a vertex.

We can summarize quite compactly our results as follows:\\
Let us introduce the function $V(x)\equiv
\sum_k \tau _k \frac{ {x}^k}{k!}$, which we call the potential
for reasons which will be clear in a moment. If all our previous
formulae are to make sense, this function should have a positive
radius of convergence.\\
Let us also define
\begin{equation}
{\mathcal F}(y,x)\equiv -1- \log y -\frac{x^2}{2}+yV(x).
\label{freemax}
\end{equation}
Then $(\bar{y},\bar{x})$ is a critical point for ${\mathcal F}$,
thanks to eqs.(\ref{1yvx},\ref{xbar}),
and $F$ is the corresponding critical value.

It is not true that these equations for $(\bar{y},\bar{x})$ always
have a single solution. It is not difficult to find examples with no
solution at all. We can interpret this by saying that in that case
there is no thermodynamic limit in our sense. More troublesome is the
case when there are several solutions. The most naive requirement
would be that the physical solution is to take the couple
$(\bar{y},\bar{x})$ that leads to the absolute maximum $F_{max}$ for
$F$ because the factor $e^{NF_{max}}$ will be the dominant
contribution to $Z$. We shall meet such a behaviour in one of the examples
of Section \ref{sec:recons}, and make a few comments there. 

For most of the paper, we shall simply assume that if there is more
than one extremum, we have picked the correct one.  

\subsection{Connected components}
\label{sec:conncomp}
 
In section \ref{sec:compdis}, we gave a formula for $Z_N$ in terms of
connected components. This formula has also an interesting limiting
form in the thermodynamic limit, but we shall wait until the next
section to derive it. For the time being, recall that $W_n$ is the sum of the
weights of connected graphs on $n$ vertices. We have shown that 
 the average number of components of
size $n$ in the random graph is 
$\frac{N!}{n!(N-n)!} W_n\frac{Z_{N-n}}{Z_N}.$

Now, we split $W_n=\sum_{l \geq 0} W_{n,l}$ as a sum of
contributions corresponding to connected graphs with $l=0,1,\cdots$
(independent) loops\footnote{Or closed circuits in the mathematical
  literature.} . If $G$ is a connected graph with $L$ (independent)
loops, $E$ edges and $V$ vertices, an old theorem of Euler says that
$L=E-V+1$ (in particular trees, i.e. connected graphs without loops,
have $E=V-1$) so $W_{n,l}$ is simply the homogeneous component of
degree $2(n+l-1)$ in the $t_k's$. If we set $t_k=\tau _k N^{-k/2}$ we
see that $W_{n,l}(t_.)=N^{1-n-l}W_{n,l}(\tau _.)$. We define
$ T_n \equiv W_{n,0}(\tau _.)$.

When $N \rightarrow \infty$ for fixed $n$ and $\tau _k$'s we find that
$\frac{N!}{n!(N-n)!} W_n \sim N \frac{T_n}{n!}$, meaning that trees
dominate. In the  thermodynamic limit, we find as before that 
$\frac{Z_{N-n}}{Z_N}(t_.)\sim
e^{-n(F+\frac{1}{2}\sum_k k\tau _k\frac{\partial F}{\partial \tau
_k})}\equiv \bar{y}^n.$

So in the thermodynamic limit, the average number of components of
size $n$ in the random graph is $N \frac{T_n}{n!} \bar{y}^n.$
The number of points in components of size $n$ in the random graph is
$$ C_n\equiv N n\frac{T_n}{n!} \bar{y}^n,$$
and the total fraction of sites occupied by finite components is 
$$ Q\equiv \sum _n n\frac{T_n}{n!} \bar{y}^n.$$
If this number is $1$, we can consistently interpret the random graph
model as a random forest model in the thermodynamical limit. 
However, if this number is $< 1$, this means that a finite fraction of
 points is not in finite components, and there is a percolation
cluster in the system.

\subsection{Tree distribution}
\label{sec:tree}

We would like to find a closed formula for the generating function
$$ T(y) \equiv \sum _n n\frac{T_n}{n!} {y}^n.$$

The first observation comes from an analogy with a baby quantum field
theory.
The asymptotic  expansion of the integral 
$$I= \frac{1}{(2\pi \hbar)^{1/2}}\int _{-\infty} ^{+\infty}
e^{(-x^2/2+yV(x))/\hbar}$$  in powers of the $\tau_k$'s 
has a useful reinterpretation. Namely 
$$I \sim \sum_G \frac{1}{A(G)} \hbar^{L(G)-C(G)}\prod _k (y\tau
_k)^{\tilde{G}_k},$$
where the sum is over Feynman
graphs\footnote{Warning : Feynman graphs are essentially general
graphs i.e.  not necessarily simple !} with an arbitrary number of
vertices, $A(G)$ is the order of the automorphism group of $G$ (for a
precise definition see e.g.  \cite{Itzyk}) and $C(G)$ the number
of connected components of $G$. Again by a factorization
argument for the weights the connected contributions exponentiate, and
$$ 
\hbar \log I \sim {\sum_{G} }^c
\frac{1}{A(G)}\hbar^{L(G)} \prod _k (y\tau _k)^{\tilde{G}_k}.
$$
In the classical (small $\hbar$) limit, on the one hand graphs with
$L(G)=0$ dominate. Though Feynman graphs are not
necessarily simple, loopless Feynman graphs are just ordinary trees.
On the other hand, $I$ can be calculated 
in the limit $\hbar$ small by the
saddle point approximation, leading to the identity between formal
power series:
\begin{equation}
T(y)\equiv {\sum_{{\rm loopless  }\, G}}^{\hskip -.3 truecm c} \hskip .3 truecm
\frac{1}{A(G)} \prod _k (y\tau _k)^{\tilde{G}_k} = S(\tilde x)
\label{Tformel}
\end{equation}
with $S(x)=-x^2/2+yV(x)=-x^2/2+y\sum_k \tau_k x^k/k!$ and $\tilde x$ 
is the formal power series of $y$ and the $\tau _k$'s for which $S$ is extremal,
$\tilde x=y\sum_k \tau_{k+1} {\tilde x}^k/k!=yV'(\tilde x)$.

Hence, the expansion of $-{\tilde x}^2/2+yV(\tilde x)$ in a formal power series of $y$ 
with $\tilde x=yV'(\tilde x)$ yields $\sum_n y^n T_n/n! $. 
From $ T(y)=-{\tilde x}^2/2+yV(\tilde x)$ and the stationnarity condition, 
we also infer that $T'(y)=V(\tilde x)$.

The expansion of $T(y)$ is convergent for small $y$ 
if $V(x)$ has a non vanishing radius of convergence. 
Note that if $\tau _1=0$, the solution $\tilde x=0$ has to be
chosen, because it leads to the correct $T(y)=\tau _0 y$ (trees on
two or more vertices have leaves, so they count $0$ if $\tau _1=0$).

We now make the general assumption that $\tau _1 \neq 0$,
and $V(x)$ has a non-vanishing radius of convergence.
Let us study the inversion of the relation $y=x/V'(x)$.
This can be obtained via the Lagrange formula.
By Cauchy's residue formula
$$ \tilde x(y)=\oint x\,\frac{(x/V'(x))'}{x/V'(x)-\tilde x/V'(\tilde x)}dx$$ 
where the $x$ contour has index $1$ with respect to $\tilde x$. 
Replacing $\tilde x/V'(\tilde x)$ by $y$, the $y$-expansion yields
$$\tilde x(y)=\sum_{n \geq 0} y^n \oint
x\,\frac{(x/V'(x))'}{(x/V'(x))^{n+1}} dx.$$
One can use integration by parts to get:
$$\tilde x(y)=\sum_{n \geq 1}\frac{y^n}{n!}
\left(\frac{d^{n-1}}{dx^{n-1}}
\left(\frac{dV}{dx}\right)^n\right)_{|x=0}.$$
This is clearly a series with non-negative coefficients.  As a
consequence, its radius of convergence is given by the first
singularity on the positive real axis, at the point $y_m=x_m/V'(x_m)$
corresponding to the unique maximal value of the concave function
$x/V'(x)$. This maximum might have two origins : either $x_m$ is a
singular point of $V$, or the derivative of $x/V'(x)$, which is
positive for small $x$, vanishes at $x_m$. This  is equivalent to
$V'(x_m)-x_mV''(x_m)=0$

The explicit form of $T_n$ can similarly  
be obtained via the Lagrange formula:
$$T(y)=\tau _0 y+\sum_{n \geq 1}\frac{y^n}{n!}
\left(\frac{d^{n-2}}{dx^{n-2}}
\left(\frac{dV}{dx}\right)^n\right)_{|x=0},$$
a classical formula which can also be proved by purely combinatorial
arguments, giving an independent argument for the fact that the
classical limit of quantum field theory is described by trees.

\subsection{Percolation}
\label{sec:perco}

The  analysis of the previous section shows that 
 the series $$\sum_n n(x/V'(x))^{n-1}\, T_n/n!$$
converges for any positive $x$ in the domain of convergence of $V$, 
and that, in this domain, its sum is equal to $V(x)$ for $x<x_m$.

For $x>x_m$ the series is still convergent. However there is a (unique)
number $x^*<x_m$ such that $x^*/V'(x^*)=x/V'(x)$ and, since
the series only involves the ratio $x/V'(x)$, we have: 
$$\sum_n n(x/V'(x))^{n-1}\, T_n/n!=V(x^*) < V(x),
\quad {\rm for}\quad x>x_m.$$

The percolation question can be rephrased as follows. Is the
relevant solution $\bar x$ of the system $ \bar{y}V'(\bar{x})=\bar{x}$,
$\bar{y}V(\bar{x})=1$ such that $\bar{x} \leq x_m$ or not? Indeed, we
know that the fraction of points in finite clusters is $\sum_n
n\frac{T_n}{n!}\bar{y}^{n}$. Substituting
$\bar{y}=\bar{x}/V'(\bar{x})$, this series sums to
$\bar{y}V(\bar{x})=1$ if $\bar{x} \leq x_m$ but to
$\bar{y}V(x^*(\bar{x}))<1$ if $\bar{x} > x_m$.  The condition $\bar{x}\,
{}_>^<\, x_m$ is equivalent to the condition
$V'(\bar{x})-\bar{x}V''(\bar{x})\,{}_>^<\, 0$. This can be
transcribed in term of the probability distribution $\{\pi_k\}$ :
$\bar{y}\bar{x}V'(\bar{x})=\sum_k k \pi_k\equiv \langle k \rangle$ and
$\bar{y}\bar{x}^2V''(\bar{x})=
\sum_k k(k-1) \pi_k\equiv \langle k(k-1) \rangle$ (we
use brackets to denote averages of the distribution $\{\pi_k\}$).

Thus, the percolation criterion is that there is a percolation cluster
in the system if and only if $\langle 2k-k^2 \rangle<0$. 
This is precisely the criterion given in ref.\cite{molloy}. 
The relative size of the giant component $Q_\infty = 1 - \sum_n 
n\frac{T_n}{n!}\bar y ^n$ is then:
\begin{equation}
Q_\infty = 1 - \bar y V(\bar x^*)
= 1 - \sum_k \pi_k (\bar x^*/\bar x)^k
\label{qperco}
\end{equation}
where $\bar x^*$ is the smallest $x$ solution of $\bar{y}=x/V'(x)$.
This is again in agreement with the result of ref.\cite{molloy}. Close
to the percolation threshold, and for a generic potential $V$, the
size of the giant component increases linearly with $\langle k^2-2k
\rangle$:
$$
Q_\infty \simeq \frac{ \langle 2k\rangle\, \langle k^2 -2k\rangle}{
\langle k(k-1)(k-2) \rangle}
$$
This formula is not valid when the probability distribution $\pi_k$
has no third moment. Then the grows of the giant component close to
the transition can exhibit a different critical behavior. We shall
give an example of this situation in the examples of section \ref{sec:recons}.

Let us analyse in more details what happens if $\bar{x}= x_m$. We know
that there is no percolation cluster. Now, if the radius of
convergence of $V$ is strictly larger than $ x_m$, close to
$y_m=x_m/V'(x_m)$, $T'(y)$ has a square root branch point. This
implies that the contribution of points in components of size $n$ in
the system decreases algebraically as $C_n \sim  N n^{-3/2}$ for
large $n$. In the physics language, this is interpreted as a critical
point and $3/2$ as a critical exponent. Note that even in this case,
the distribution $\{\pi_k\}$ is still decreasing at least
exponentially at large $k$. 

To observe other critical points, with
different critical exponents, the radius of convergence of $V$ has to
be exactly $\bar{x}= x_m$, which requires some fine tuning.
In that case, both $C_n$ and $\pi_k$ decrease algebraically. 
Assume that  $V$ has a leading singularity at $\bar x=x_m$
locally of the form $(\bar x - x)^\gamma$, with $\gamma>2$ to ensure
the existence of $\langle k \rangle$ and $\langle k^2 \rangle$. 
Generically, $y-y_m$ is linear
in $x-x_m$ so that $yT'(y)=V(x)$ has a leading singularity of the form
$(y-y_m)^\gamma$ and both $\pi_k$ and $C_k/N=kT_k\bar y^k/k!$ decrease 
as $k^{-\gamma -1}$.
We shall give an example below.

If there is no percolation cluster, we can treat the large $N$ limit
from another point of view. We start from eq.(\ref{superzn}) and in the contour
integral giving $Z_N$, we change variables and replace $z \rightarrow
Nz$, leading to
$$ Z_N = \frac{N!}{N^N}\oint \frac{dz}{z}z^{-N} e^{N\sum_{l \geq
 0}N^{-l}\sum_{n \geq 1} \frac{z^n}{n!}W_{n,l}(\tau _.)}.$$
For fixed $n$, the connected graphs with loops ($l \geq 1$) are
suppressed by inverse powers of $N$. However, in the sum over the size
of connected components, terms up to $n=N$ make a contribution to the
contour integral, and it might happen that for large $n$ and $N$
related by some condition connected components of size $n$ of with
loops make a finite contributions to $\frac{1}{N}\log Z_N$. However,
if there is no percolation cluster, we may safely neglect $l \geq 1$
and get an
accurate approximation to the leading exponential behavior of
$Z_N$ in the large $N$ limit.

Under appropriate conditions, the contour
integral for $Z_N$ can be deformed to pass through a dominant saddle
point.  Then the free energy is given by the
saddle point approximation. We see that $Z_N \sim e^{NF(\tau _.)}$ with
$ F(\tau _.)=-1-\log \bar z +T(\bar z)$,
$\bar z$ being the the saddle point maximizing 
$ -1-\log z +T(z)$. This equation is what one gets from eq.(\ref{freemax})
 when $\bar y=\bar z$ and $\bar x$ is seen
as a function of $\bar y=\bar x /V'(\bar x)$. This gives yet another proof
of the dominance of trees and the Lagrange inversion formula.

\subsection{Conditional degree distributions}
We now present formulas for the degree distributions,
denoted $\pi_k^{(n)}$,
for vertices within clusters of size $n$. 
We are particularly interested in the degree distribution
$ \pi_k^{(\infty)}$ in the giant component when it exists.

 From the last formula of Section \ref{sec:compdis}, the average number 
of vertices of degree $k$ belonging to a component of size $n$ is:
$$
C_n(k)= \frac{N!}{n!(N-n)!}\, (t_k\frac{\partial W_n}{\partial t_k})\,
\frac{Z_{N-n}}{Z_N}
$$
In the thermodynamic limit, $t_k=N^{-k/2}\tau_k$, $N\to \infty$,
this becomes 
$$
\frac{C_n(k)}{N} =\frac{{\bar y}^n}{n!}(\tau_k\frac{\partial T_n}{\partial \tau_k}).
$$
By definition, the degree distribution within components of size $n$
is $C_n(k)$ divided by the average number of points in components
of size $n$ so that $\pi_k^{(n)}=C_n(k)/C_n$,
 with $C_n/N=nT_n{\bar y}^n/n!$ in the thermodynamic limit.
Hence:
$$\pi_k^{(n)}=\frac{1}{n}\,\tau_k \frac{\partial\log T_n}{\partial \tau_k}.
$$
Notice that these distributions are normalized, $\sum_k \pi_k^{(n)}=1$,
since the $T_n$'s are homogeneous polynomials in the $\tau_i$ of degree $n$
if each $\tau_i$ is assigned degree one.

Assume now that the percolation criterion is satisfied so that a giant
component exists. 
The number of vertices of degree $k$ in the giant component are:
$C_\infty(k)= N\pi_{k} - \sum_n C_n(k)$. 
In the thermodynamic limit,
$$
C_\infty(k)/N=\pi_k - \sum_n(\tau_k\partial_{\tau_k}T_n)\, \frac{ {\bar y}^n}{n!}
=\pi_k - (\tau_k \partial_{\tau_k} T)(\bar y)
$$
But $T(y)=-\tilde x/2+yV(\tilde x )$ with the extremum condition 
$\tilde x=yV'(\tilde x)$ so that 
$(\partial_{\tau_k} T)(y)=y(\tilde x)^k/k!$. 
Using $\pi_k=\bar y\tau_k \frac{\bar x^k}{k!}$, we get,
with $\bar x^*$ defined as in Section \ref{sec:perco}:
$$
\frac{C_\infty(k)}{N}=\pi_k\Big(1 - \Big(\frac{\bar x^*}{\bar x}\Big)^k\Big)
$$
or equivalently,
\begin{equation}
\pi^{(\infty)}_k = \frac{\pi_k}{Q_\infty}\,
\Big(1 - \Big(\frac{\bar x^*}{\bar x}\Big)^k\Big)
\label{pinfty}
\end{equation} 
since $\pi^{(\infty)}_k$ is the ratio between $C_\infty(k)$ and the
number of points in the giant cluster, which is $NQ_\infty$.  As it
should, $\pi^{(\infty)}_k$ is correctly normalized:
$\sum_k\pi^{(\infty)}_k=1$, and vanishes at $k=0$ (there is no
isolated vertex in the giant component). There is a crossover value
$k_c=\log(\bar x/\bar x^*)^{-1}$ above which
$\pi^{(\infty)}_k$ is exponentially close to $\pi_k/Q_\infty$. 
Close to the transition the ratio $\pi^{(\infty)}_k/\pi_k$ goes
to $k/\langle k \rangle $.
 The formula for $\pi^{(\infty)}_k$ has a simple probabilistic
interpretation : as it is the conditional probability that a vertex
has $k$ neighbors given that it is in the giant component, it can be
written as the quotient of $\pi_{k,\infty}$, the probability to have
$k$ neighbors {\it and} be in the percolation cluster, by $Q_\infty$.
We read from eq.(\ref{pinfty}) that 
$\pi_{k,\infty}=\pi_k - \pi_k\Big(\frac{\bar x^*}{\bar x}\Big)^k$. 
Hence $\pi_k\Big(\frac{\bar x^*} {\bar
x}\Big)^k$ is the probability for a vertex to have $k$ neighbors and
to be in a finite component. This suggests that
 when a new point is added to the graph,
the probability that it connects to $k$ other vertices none of them in
the giant component is $\pi_k\Big(\frac{\bar x^*} {\bar
x}\Big)^k$: for each new edge, the penalty for avoiding the giant
component is $\frac{\bar x^*}{\bar x}$

\subsection{Reconstruction, with examples}
\label{sec:recons}
The maximal entropy graph distribution can be reconstructed form the
data of the degree distribution $\pi_k$, $\sum_k\pi_k=1$. We set
$H(s)=\sum_k\pi_k s^k$. 

Given $\pi_k$, $\bar{x}$ is defined as the positive square root of
$\langle k \rangle=\sum_k k\pi_k$, and $\bar{y}\tau_k$ as $\pi_k\,
k!/\bar{x}^k$. This yields $\bar{y}V(x)=\sum_k\pi_k
(x/\bar{x})^k=H(x/\bar{x})$.  The coefficient $\bar{y}$ appears then
as a normalization factor which may be choosen at will, eg. we could
set $\bar{y}=1$. 

The tree distribution $T(y)=\sum_n T_ny^n/n!$ is then reconstructed,
as a formal series, from $T'(y)=V(\tilde x)$ with $\tilde x=yV'(\tilde x)$. 
It is clear that   
$T_n\bar{y}^n$ is independent of the choosen normalization for $\bar{y}$. 

The fraction of site occupied by the finite size components
is $Q= \bar{y}T'(\bar{y})$.
By construction, $\bar{x}$ is solution to $\bar x/V'(\bar x)=\bar{y}$.
The giant component exists when there are two solutions to
the above equation in the interval $[0,\bar{x}]$. 
We denote by $\bar x^*$ the smallest of them. 
The fraction of site occupied by the giant component is
$Q_\infty = 1 - \bar{y}V(\bar x^*)$. Equivalently, one can look for a
solution $0< s^* <1$ to the equation $H'(s)=sH'(1)$, and if there is
one, $Q_\infty=1-H(s^*)$.   

\vspace{.3cm}

Let us illustrate this reconstruction on a few simple examples.
\begin{enumerate}
\item Poissonian degree distribution: $\pi_k=e^{-\alpha}\, \alpha^k/k!$.
This is the Erd\"os--Renyi model. We have $\bar{x}=\alpha^{1/2}$,
and $V(x)=\exp x\bar{x}$, choosing $\bar{y}=e^{-\alpha}$.
The tree distribution is $T'(y)=\exp\tilde x$ with $\tilde x
e^{-\tilde x}=\alpha y$. The giant component exists for $\alpha>1$
when the equation $\tilde xe^{-\tilde x}=\alpha e^{-\alpha}$ admits
two solutions ${\alpha}^*$ and $\alpha$ with
${\alpha}^*<1<\alpha$.
Its relative size is $Q_\infty=1-e^{-\alpha}T'(e^{-\alpha})=
1-{\alpha}^*/\alpha$.

\item Geometric degree distribution: $\pi_k=(1-p)p^k$. Then 
$\bar x^2= p/(1-p)$ and $\bar y V(x)=1/(1+\bar x^2-x\bar x)$.
The extremum relation $x=yV'(x)$ is a cubic equation:
$\bar y x(1+\bar x^2-x\bar x)^2=y\bar x$. The percolation transition is
at $\bar x^2=1/2$ $(p=1/3)$ and the relative size of the giant
component is
$Q_\infty = 1 -(\bar x^*/\bar x)^{1/2}$ with 
$\bar x \bar x^*=\frac{1}{2}(\bar x^2 +2 - \bar x \sqrt{4+\bar x^2})$.

This example confronts us with the ambiguity problem alluded to a long
time ago. 

Changing $p$ into $1-p$ leads to replace $\bar x$ by $1/ \bar x$. 
This changes $\bar y V(x)=1/(1+\bar x^2-x\bar x)$ 
into itself up to an irrelevant 
multiplicative factor. To state things in a slightly different way,
the extremum conditions $yV'(x)=x$ and $yV(x)=1$ have two solutions,
and one leads to the geometric distribution with parameter $p$ and the
other one to the geometric distribution with parameter $1-p$. Of
course, the real result of the computation of $Z_N$ will make a
definite choice. The criterion of the maximum for $F$ leads to choose
$\inf (p,1-p)$, and this is also consistent with continuity starting
from $p=0$, a random graph made of isolated points.

However, the formulas obtained before for, say, the size of the giant
component, coincide with the ones from ref.\cite{molloy} even for
$p > 1/2$. This situation requires clarification. Maybe this is the
point when the canonical and the microcanonical approaches finally
diverge and stop being equivalent.  

\item An example of a scale free distribution: $H(s)=\sum \pi_k s^k \equiv
\tau_0 +\tau_1 s +\tau_2 s^2/2+\tau(1-\beta s+\beta(\beta-1)
s^2/2-(1-s)^{\beta})$, where $2<\beta<3$ and $\tau_0,\tau_1,\tau_2$ 
and $\tau$ are nonnegative parameters subject to the condition
$\tau_0 +\tau_1 s +\tau_2 s^2/2+\tau(\beta-1)(\beta-2)/2=1$ to ensure
that the $\pi_k$'s are correctly normalized. The $\pi_k$'s decrease
like $\pi_k \sim \frac{-\tau}{\Gamma (-\beta)}k^{-\beta -1}$.

Then $\bar x
^2=H'(s=1)=\tau_1  +\tau_2 +\tau \beta (\beta-2)$ from which the
potential $\bar y V(x)$ is recovered as usual.  

There is a percolation cluster if and only if the equation
$H'(s)=sH'(1)$ has a solution $s^* <1$. So, we look for the solutions
of $(1-s)(\tau \beta -\tau_1)=\tau \beta (1-s)^{\beta -1}$. If
$\langle k^2-2k\rangle=\tau  \beta-\tau_1$ is negative, there is no
percolation cluster, but if it is positive,
$1-s^*=(1-\frac{\tau_1}{\tau \beta})^{1/(\beta-2)}$. The size of the
giant component is $1-H(s^*)\sim (1-s^*)H'(1)$, and 
$$Q_\infty \sim \langle k^2-2k\rangle^{1/(\beta-2)}$$
close to the threshold. This is an example when the
growth of the giant component close to the threshold is nonlinear as a
function of $\langle k^2-2k\rangle$.

The number of points in components of size $k$ is reconstructed from 
$T'(y)=V(x)$ with $x=yV'(x)$. Below the threshold, this leads to a
singularity $T'_{sing}\sim (\bar y -y)^{\beta}$, which implies that
$C_n \sim n^{-\beta -1}$. Above the threshold, the radius of
convergence $r$ of $T$ is larger than $ \bar y$, 
leading to $C_n$'s that decrease
exponentially as $C_n \sim n^{-3/2} (\bar y/r)^n$. 
\end{enumerate}

 \section{The case of oriented graphs}
\label{sec:orient}

It is not difficult to modify the previous arguments to deal with
maximum entropy oriented graphs with given ``in-out'' degree
distributions. We give the percolation
criterion, omitting all details. 

The first result is that for such models, each vertex
with $k$ outgoing and $l$ incoming vertices contributes a fixed
multiplicative factor, say $t_{k,l}$, to the weight of a graph. 
The generalization of the recursion formula for $Z_N$ reads
$$
Z_N(t_{i,j}) = \oint \frac{d \omega_+}{\omega_+}  
\frac{d \omega_-}{\omega_-}\sum_{k,l} t_{k,l}
\omega_+ ^{-k}\omega_- ^{-l}Z_{N-1}(t_{i,j}+\omega_+
t_{i,j+1}+\omega_-t_{i+1,j}).
$$
The large $N$ finite connectivity limit is obtained by letting $N
\rightarrow \infty$ while keeping $\tau _{k,l}=t_{k,l}N^{(k+l)/2}$
fixed. Defining 
$$V(x_+,x_-)\equiv \sum_{k,l} \tau _{k,l} \frac{
x_+^k}{k!}\frac{ x_-^l}{l!},$$
 a straightforward adaptation of
the argument in section \ref{sec:genal} leads to the fact that the
free energy $F$ is the value of
$$
{\mathcal F}(y,x)\equiv -1- \log y -x_+x_-+yV(x_+,x_-).
$$
at the point $(\bar{y},\bar{x}_+,\bar{x}_-)$ where it is maximum :
\begin{equation}
\bar{x_+}=\bar{y}(\partial _{x_-} V)(\bar{x}_+,\bar{x}_-),
\quad \bar{x_-}=\bar{y}(\partial _{x_+} V)(\bar{x}_+,\bar{x}_-),
\quad 1=\bar{y}V(\bar{x}_+\bar{x}_-).
\label{orientvvv}
\end{equation}

The analysis of the first two equations is a bit more involved than
the analysis of the single implicit equation for the oriented case.
We can view the pair of equations $x_+=y\partial _{x_-} V$ and
$x_-=y\partial _{x_+} V$ in the following way. It defines a function
$y$ over the curve $\mathcal C$ in the positive quadrant of 
the $(x_+,x_-)$ plane given by
$x_-\partial _{x_-} V=x_+\partial _{x_+} V$. This curve is smooth as
long as $V$ is well-defined. For instance, one can take $ x \equiv
x_+\partial _{x_+} V=x_-\partial _{x_-} V$ as an analytic parameter on
it. Then $y$ is a smooth convex function of $x$, and all properties of
the non-oriented case are true for $y(x)$ : $y$ is a good analytic
parameter on $\mathcal C$ for small $y$, but there is a singularity if
the convex function $y(x)$ has a maximum. To be more explicit, taking
differentials we see that
$$
\left(\begin{array}{c} dx/x \\ dx/x \end{array} \right)
=\left(\begin{array}{cc} 1+x_+ \partial _{x_+}^2 V/\partial _{x_+}V &
x_- \partial _{x_+}\partial _{x_-} V/\partial _{x_+}V \\ x_+ \partial
_{x_-}\partial _{x_+} V/\partial _{x_-}V &  1+x_- \partial _{x_-}^2
V/\partial _{x_-}V\end{array} \right) \left(\begin{array}{c} dx_+/x_+
\\ dx_-/x_- \end{array} \right), 
$$
and 
$$
\left(\begin{array}{c} dy/y \\ dy/y \end{array} \right)
=\left(\begin{array}{cc} 1-x_+\partial _{x_-}\partial _{x_+}
V/\partial _{x_-}V & -x_- \partial _{x_-}^2 V/\partial _{x_-}V \\
-x_+ \partial _{x_+}^2 V/\partial _{x_+}V & 1-x_- \partial
_{x_+}\partial _{x_-} V/\partial _{x_+}V\end{array} \right)
\left(\begin{array}{c} dx_+/x_+ \\ dx_-/x_- \end{array} \right). 
$$
A simple computation shows that the determinant of the $2$ by $2$
matrix in the first relation is always strictly positive, but that 
the determinant of the $2$ by $2$ matrix in the second relation is
positive for small $y$ but can change sign. This happens if $y'(x)$
vanishes. So the discussion of the non-oriented case carries over word
for word. It is consistent to write $\bar{y}=y(\bar{x}),
\bar{x}_\pm=x_\pm(\bar{x})$ and $\bar x^*(\bar{x})$ 
for the smallest $x$ such that $y(\bar{x})=y(x)$.
We set $\bar x_\pm^*\equiv x_\pm(\bar x^*)$. 
There is a percolation cluster if and only if
the second determinant is $<0$ at
$(\bar{y},\bar{x}_+,\bar{x}_-)$. This is equivalent to
$$
(\partial _{x_-}V-x_+\partial _{x_-}\partial _{x_+}V)(\partial
_{x_+}V-x_-\partial _{x_+}\partial _{x_-}V)-x_+x_- \partial _{x_-}^2 V
\partial _{x_+}^2 V$$ 
being $<0$ at that point. 
Then the fraction of sites in the percolation cluster is 
$$ Q_{\infty}=1-\bar{y}V(\bar x_+^*,\bar x_-^*).$$

To get the percolation criterion, we just have to rephrase the
vanishing of the determinant in terms of the probability distribution
$\pi _{k_+,k_-}$ that a vertex of the random graph has $k_+$ outgoing and
$k_-$ incoming vertices. The explicit formula is
\begin{equation}
\pi _{k_+,k_-} = \bar{y} \tau _{k_+,k_-}
\frac{{\bar {x}_+}^{k_+}}{k_+!}\frac{{\bar {x}_-}^{k_-}}{k_-!}.
\label{doublepi}
\end{equation}
By construction $\sum_{k,l}  k\pi_{k,l} 
= \sum_{k,l} l\pi _{k,l}\equiv\langle k\rangle $ since any graph has
the same number of outgoing and incoming edges.
The parameters $\bar{x}_+$ and $\bar{x}_-$ are constrained by the
relation 
$$ \bar{x}_+\bar{x}_-=\langle k\rangle $$

The percolation criterion  reads: 
\begin{equation}
(\langle k\rangle-\langle k_+k_-\rangle)^2 
-\langle k_+^2-k_+\rangle\langle k_-^2-k_-\rangle <0.
\label{cricri}
\end{equation}

Given the distribution $\pi_{k,l}$, the potential $V$ can be 
reconstructed via eq.(\ref{doublepi}). As in the oriented case,
the parameter $\bar y$ is
an arbitrary normalization factor. The product of the
parameters $\bar x_\pm$ is determined by $\bar{x}_+\bar{x}_-=\langle k\rangle$.
The ratio $\bar x_+/\bar x_-$ can be choosen at will since there is
a natural  invariance in eq.(\ref{orientvvv}). Namely if 
$\bar x_\pm$ solves the extremum condition for the  potential 
$V(x_+,x_-)$, so does $\hat x_\pm=\lambda^{\pm 1} \bar x_\pm$
for the potential $W(x_+,x_-)=V(\lambda x_+,\lambda^{-1} x_-)$
and leaves $\pi_{k,l}$ invariant. 
Again this finds its origin in the fact that any graph has
the same number of outgoing and incoming edges.

\vskip 1.5 truecm

\textbf{Acknowledgments}: We thank the E. Schr\"odinger Institut in
Vienna for hospitality during the completion of this work.

\end{document}